
\magnification=\magstep1
\baselineskip = 18 true pt
\hsize =  17 true  cm
\vsize =  23 true cm
\centerline {{\bf PEAKONS, R-MATRIX AND TODA LATTICE}}
\vskip 1 true cm
\centerline {$ O.Ragnisco^{(a)}, M.Bruschi^{(b)}$ }
\centerline {{\it Istituto Nazionale di Fisica Nucleare, Sezione di Roma,
P.le A.Moro 2, 00185 Roma,Italia} }

\centerline {{\it (a) Dipartimento di Fisica E.Amaldi, III Universit\`a
 di Roma }}
\centerline {{\it (b) Dipartimento di Fisica, Universit\`a di Roma
``La Sapienza" }}
                        \vskip 1 true cm

\vskip 2 true cm
\vskip 1 true cm
\noindent {\bf Abstract}

\noindent
The integrability of a family of hamiltonian systems, describing
in a particular case the motion
of N ``peakons" (special solutions of the so-called Camassa-Holm
equation)
is established
in the framework of the $r$-matrix approach, starting from its
Lax representation.
In the general case, the $r$-matrix is a dynamical one and has an
interesting though complicated structure. However, for a particular
choice of the relevant parameters in the hamiltonian (the one
corresponding to the pure ``peakons" case), the $r$-matrix
becomes essentially constant, and reduces to the one pertaining
to the finite (non-periodic) Toda lattice. Intriguing consequences
of such  property are discussed and an integrable time discretisation
is derived.
\vfil\eject

\noindent {\bf 1. Introduction}

Interest in integrable Nonlinear Partial Differential Equations
(NLPDEs) modelling physical phenomena and possessing ``exhotic"
localized solutions has been revived in recent years thanks to the work of
Camassa and Holm [1].

Namely, in [1] an integrable NLPDE, included as a special case in the general
scheme provided by Fokas and Fuchssteiner in 1981 [2], has been
rederived in the context of water waves, and it has been shown to
possess, under suitable assumptions, a new kind of localized solutions,
denoted as ``peakons".

On the other hand, a fruitful research activity has been in the past
devoted to describe the time behaviour of the variables characterizing
classes of special solutions of integrable NLPDEs in terms of
integrable hamiltonian systems with finitely many degrees of freedom
[3 - 7].
Quite naturally then, a similar approach has been pursued to investigate
the N-``peakons" dynamics, by different authors with different perspectives,
and results have been remarkable [1, 8, 9].
In particular, in ref.[9] it has been proven the complete integrability of
the hamiltonian system defined, in canonical coordinates, by the
following Hamilton's function:

$${\cal{H}}={1\over 2}\sum_{j=1}^N \sum_{k=1}^N p_jp_k(\lambda +
 \mu {\rm{cosh}}[\nu (q_j-q_k)] + \mu^\prime {\rm{sinh}}[\nu
\vert q_j-q_k \vert])
\eqno(1.1)$$

\noindent
which reduces, in the special case $\mu^2=(\mu^\prime)^2$, to the one
describing the motion of N ``peakons" of the Camassa-Holm equation.

In this paper, the complete integrability  of the hamiltonian $(1.1)$ will
be proven in  the framework
the classical $r$-matrix approach
[10, 11]. In this way, we will not  only succeed in
casting the system under scrutiny in a general unifying scheme, but we will
also be able to uncover the basic features that render the pure
``Peakons-Lattice" case $\mu^2=(\mu^\prime)^2$ (hereby denoted as PL)
quite peculiar with respect to the general case.

In Section 2,
starting form the Lax representation constructed in [8], we derive
the $r$-matrix for the general case (that will be in the following  denoted
as GL), which turns out to be of dynamical
type; its elegant structure allows one to give a neat characterisation
of the Lax pair associated with any of the flows commuting with the
${\cal{H}}$ flow. In Section 3, we restrict our considerations to
PL, and observe first of all that its $r$-matrix is essentially constant,
and is identical with the one pertaining to the finite Toda lattice;
the two systems are thus endowed with the same Lie-Poisson structure.
Guided by the above finding, we prove that the PL dynamics is
just, in different coordinates, one of the compatible dynamics of the
Toda lattice. Moreover, by using the factorisation (QR) algorithm [12],
we construct an integrable time discretisation (Backlund transformation)
for PL. Some possible extensions are outlined in Section 4.
\vskip 1 true cm
\noindent {\bf{2.~The $r$-matrix for the General Lattice}}

First of all. we write down explicitly the Hamilton's equations
associated with $(1.1)$:
$$\dot p_j=-{\partial  {\cal{H}}\over \partial q_j}=~p_j\sum_k p_k
\{-\mu{\rm{sinh}}[\nu(q_j-q_k)]~+~\mu^\prime {\rm{sign}}(q_j-q_k)
{\rm{cosh}}[\nu(q_j-q_k)]\}
\eqno(2.1a)$$

$$\dot q_j={\partial  {\cal{H}}\over \partial p_j}=~\sum_k p_k
\{\lambda+\mu{\rm{cosh}}[\nu(q_j-q_k)]~+~\mu^\prime {\rm{sinh}}
[\nu\vert q_j-q_k \vert]\}
\eqno(2.1b)$$

\noindent
We  shall assume both $\mu$ and $\mu^\prime$ not equal to $0$. Indeed, the
behaviour of the system in these special cases has been already investigated
by F.Calogero in a previous paper [13], and consequently will not
be considered here. With no restriction, we can then rescale to $1$
one of the parameters, say $\mu$, so that PL corresponds to $\mu^\prime=\pm 1$.

Following  basically the notations used in [8], we write the Lax representation
for (2.1) in the form:

$$\dot L~=~[L,M]\eqno(2.2a)$$

\noindent
where $L,M$ are $N\times N$ matrices, whose entries read:

$$L_{jk}={\sqrt{p_j p_k}}\alpha (q_j -q_k)\eqno(2.2b)$$

$$M_{jk}=-2\kappa{\sqrt{p_j p_k}}\alpha^\prime (q_j -q_k)\eqno(2.2c)$$

\noindent
In (2.2c) a prime denoted differentiation with respect to the argument, and
the function $\alpha (x)$ is given by:

$$\alpha (x)~=~{\rm{cosh}}{\nu x \over 2}~+~\rho {\rm{sinh}}{\nu \vert x \vert
 \over 2}
\eqno(2.3)$$

\noindent
with:

$$\mu^\prime~=~-{2\rho \over 1+\rho^2};~~\kappa~=~{2 \over 1+\rho^2}$$

\noindent
The  hamiltonian (1.1) is then the following simple function of
the invariants of the matrix $L$:

$${\cal{H}}=~-\kappa~tr L^2~+~(\lambda+\kappa+1)(tr L)^2\eqno(2.4)$$

\noindent
As the term $(tr L)^2$ merely generates on overall translation of the
system, one can assume $\lambda+\kappa+1=0$ with no essential lost
of generality.

If  we rewrite for a moment the Hamiltonian (1.1) in the form:

$${\cal{H}}=\sum_{j,k} p_jp_k f(q_j-q_k),$$
 \noindent
with
$$f(x)=\lambda+1+\kappa[1-\alpha^2(x)]$$

\noindent
we readily  discover two important features of the
dynamics, which merely follow from the even nature of the function
$f(x)$, namely:

{\it{$~$i)    The $N$  coordinates $q_j$ keep their own initial
ordering ({\rm{see also [1]}}), as we have:

$${\partial \over \partial t} {\rm{sign}}(q_j-q_k)=0$$
\noindent
Hence, we  can assume   ``natural ordering": $q_1\le q_2\le \cdots \
q_N$. Physically, we would say that the particles do not cross each other.

ii) The  conjugated momenta  $p_j$ keep their own initial sign,
as we have:
$${\partial \over \partial t} {\rm{sign}}p_j=0$$
\noindent
In particular, there are no ``dynamical ambiguities" in the  expression
 ${\sqrt{p_jp_k}}$ occurring in formulas }}(2.2).

\noindent
We  further observe that Newton equations can be readily derived for the
flow generated by (1.1). The most instructive  procedure is
to introduce a Lagrangian through a Legendre transformation:

$${\cal{L}}(q,\dot q)=~ -{\cal{H}}(p,q)+\sum_k p_k\dot q_k={\cal{H}}(p,q)=
{1\over 2}\sum_{jk}\dot q_j \dot q_k G_{jk}(q)\eqno(2.5)$$

\noindent
$G(q)$ being the inverse of the matrix $(F(q))_{jk}\equiv f(q_j-q_k)$.
{}From (2.5) is natural to interpret the dynamics under scrutiny  as a
geodesic
motion on a manifold characterized by the metric tensor $G(q)$ [1].

We now turn to the evaluation of the $r-$matrix, which is well known
to be a basic tool in the theory of integrable systems [10, 11].
As it has been shown for the first time in [11], whenever
a hamiltonian system
is associated with a Lax matrix whose eigenvalues - {\it{assumed to be simple,
at least locally}} -
are in involution for a
given Poisson bracket, it enjoys an $r$-matrix representation of the form:

$$\{L_1,L_2\}~=~[r_{12},L_1]-[r_{21},L_2]\eqno(2.6)$$

\noindent
In (2.6) we have used the standard notations:

$$L_1\equiv L\otimes {\bf{1}}~=~\sum_i L_i e^i\otimes {\bf{1}}$$

$$L_2\equiv{\bf{1}}\otimes L~=~\sum_iL_i {\bf{1}}\otimes e^i$$

$$r_{12}=\sum_{i,k}r_{ik} e^i\otimes e^k~~~~~r_{21}=
\sum_{i,k}r_{ki} e^i\otimes e^k$$

\noindent
where $\{e^i\}_{i=1}^M$ is a basis for the matrix Lie-algebra ${\cal{G}}$
 which $L$ belongs
to, and, in general, the coefficients $r_{ik}$ will be functions on the phase
space ({\it{dynamical r-matrix}}).

Conversely, whenever an element $r\in {\cal{G}}\otimes{\cal{G}}$
exists such that formula (2.6) holds, the eigenvalues of $L$ and in
general its invariant functions are in involution.

\noindent
We recall  that the trace form on $\cal G$
allows one to identify $\cal G$ with its dual (the space of linear functions
on $\cal G$), and to consider the $r-$matrix as an endomorphism $R$ on $\cal
G$,
rather than as an element of $\cal G \otimes \cal G$, so that
in such a ``dual" picture,
eq.(2.6) induces the following Lie-Poisson bracket between two functions on
$\cal G$:

$$\{f,g\}_L~=~(L,[df,dg]_R)~\eqno(2.7a)$$

\noindent
with

$$[X,Y]_R=[X,R(Y)]+[R(X),Y]~~~(X,Y\in \cal G)\eqno(2.7b)$$

\noindent
and:

$$R(X)~=~\sum_{j,k}r_{jk}e^{(i)}(e^{(k)},X)\eqno(2.7c)$$

In our case, everything is very  simple. The algebra ${\cal{G}}$ is just
$gl(N,\bf R )$, so that we work with the  standard (Weyl) basis :

$$(E_{jk})_{lm}=\delta_{jl}\delta_{km}$$

\noindent
and we write:
$$\{L_1,L_2\}=\sum_{jk;lm}\{L_{jk},L_{lm}\}E_{jk}\otimes E_{lm}\eqno(2.8a)$$
$$r_{12}=\sum_{jk;lm}r_{jk;lm}E_{jk}\otimes E_{lm}\eqno(2.8b)$$

\noindent
The  basic ingredient in  the evaluation
of  the $r$-matrix is  the {\it{fundamental Poisson bracket}}
$\{L_{jk},L_{lm}\}$, whose expression in our case reads:

$$\{L_{jk},L_{lm}\}={1\over 4}[\delta_{jl}{\sqrt{p_kp_m}}(\alpha_{mj}
\alpha_{jk})^\prime + \delta_{jm}{\sqrt{p_kp_l}}(\alpha_{lj}
\alpha_{jk})^\prime $$
$$ - \delta_{kl}{\sqrt{p_jp_m}}(\alpha_{mk}
\alpha_{kj})^\prime -  \delta_{km}{\sqrt{p_jp_l}}(\alpha_{lk}
\alpha_{kj})^\prime]\eqno(2.9)$$

\noindent
In formula (2.9), a prime denotes again differentiation, and we have  used
the shorthand notation $\alpha_{jk}\equiv \alpha (q_j-q_k)$;
in the derivation of (2.9) a crucial  role  is played by the formula:

$$({\partial \over \partial x}+{\partial \over \partial y})
\alpha(x)\alpha(y)~=~
{\nu \over 4}[(1-(\mu^\prime)^2){\rm{sinh}}[\nu (x+y)]~-2\mu^\prime
\alpha(x+y)({\rm{sign}}x+{\rm{sign}}y)]\eqno(2.10)$$

Looking carefully at (2.10) we readily understand that there are two
essentially  different cases, the PL case $(\rho^2=1\to (\mu^\prime)^2=1)$
and the GL case $(\rho^2\ne 1)$.

A  comparison of formulas (2.6) and (2.9) yields the following expression
for the $r-$matrix:

$$r_{12}~=~-{\nu \over 8}\rho\sum_{jk}{\rm{sign}}(q_j-q_k)E_{jk}\otimes
(E_{jk}+E_{kj})~$$
$$+~{\nu \over  4}(1-\rho^2)\sum_{jkl}S_{kl}(E_{jk}-E_{kj})
\otimes (E_{jl}+E_{lj}),\eqno(2.11)$$

\noindent
where the {\it{symmetric}} matrix $S$ is determined, up to functions of $L$,
by the linear system:

$$[S,L]_{jk}~=~{\rm{sinh}}[{\nu \over 2}(q_j-q_k)]:= A_{jk}\eqno(2.12)$$

\noindent
It is instructive to consider the dual picture as well, where the  endomorphism
$R(X)$ (2.7c) reads:

$$R(X)={\nu \over 4}(\rho [(X^{(s)})_- - (X^{(s)})_+]~+~
2(1-\rho^2)[S,X^{(s)}]).\eqno(2.13)$$

\noindent
In (2.13) by $X^{(s)}$ we have  denoted  the symmetric part of the matrix $X$,
and  the  suffix $+(-)$ stands for the strictly upper (strictly lower)
triangular part. To write down (2.13) we took advantage  of the natural
ordering, replacing ${\rm{sign}}(q_j-q_k)$ by ${\rm{sign}}(j - k)$.

It is now a  straightforward task to write down  the Lax representation
for the  commuting hamiltonian flows  generated by the invariant  functions
${\cal{F}}^{(k)}\equiv {1\over k +1}tr L^{k+1}$. In fact we have:

$${\partial L \over \partial t_k}\equiv \{{\cal{F}}^{(k)},L\}=[L,M^{(k)}]
\eqno(2.14a),$$

\noindent
with:
$$M^{(k)}~=~R(L^k)={\nu\over 8}[\rho((L^k)_- -(L^k)_+)~+~(\rho^2-1)
\sum_{j=0}^{k-1}L^jAL^{k-1-j}\eqno(2.14b)$$

\noindent
Note  that a remarkable simplification in (2.14a) can be achieved thanks to
the identity

$$[L, \sum_{j=0}^{k-1}L^jAL^{k-1-j}]~=~[L^k,A].$$

\vskip 1 true cm
\noindent
{\bf{The PL case and its connection with the Toda lattice}}

As we have already noticed, the PL case corresponds to the
special choice $\rho^2=1$, and is thus endowed with the $r$-matrix:

$$r_{12}~=~-{\nu \over 8}\rho\sum_{jk}{\rm{sign}}(q_j-q_k)E_{jk}\otimes
(E_{jk}+E_{kj})~$$

\noindent
or, in the dual picture:

$$R_{PL}(X)={\nu \over 4}(\rho [(X^{(s)})_- - (X^{(s)})_+]$$

\noindent
which is just one  of the $r$-matrices that can be associated with the
finite nonperiodic Toda lattice [14].
Such striking coincidence strongly suggests to give a closer look to the PL
case, to uncover  possible further connections with the Toda system.

To this aim, we first notice that the inverse of the $L$ matrix (2.2b) is
in this case easily computable.

\noindent
Clearly, the problem reduces to the evaluation of the inverse of
the matrix

$$C_{jk}\equiv \alpha_{jk}~ =
{}~{\rm{exp}}(\pm {\nu \over 2} \vert q_j-q_k \vert)\eqno(3.1a).$$

\noindent
If we introduce the quantities:

$$e_j\equiv {\rm{exp}}(\pm {\nu \over 2} (q_{j+1}-q_j)\eqno(3.1b)$$

\noindent
and take into account the natural ordering, we readily see that:

$$C_{j,j+r}~=~e_je_{j+1}...e_{j+r-1}\eqno(3.1c)$$

\noindent
so that  the matrix $C$ can be explicitly inverted yielding a tridiagonal
Jacobi matrix. The final result is the  following:

$$L^{-1}={\pmatrix{B_1& A_1  & 0    &\ldots\cr
                   A_1& B_2  &A_2   &\ldots\cr
                    0 &\ddots&\ddots&\ddots\cr
               \vdots &\ldots&A_{N-1}&B_N\cr}}\eqno(3.2)$$

\noindent
where:

$$A_j=-{1\over p_jp_{j+1}}{e_j\over 1-e_j^2}~;
{}~~B_j={1\over p_j}{1-e_{j-1}^2e_j^2\over (1-e_{j-1}^2)(1-e_j^2)}\eqno(3.3)$$

\noindent
 with $e_0=e_N=0$.

\noindent
The isospectral flows of $L^{-1}$ are of course characterized by the dynamical
equations:

$${\partial L^{-1} \over \partial t_k}~=~[L^{-1},R_{PL}(L^k)]$$
\noindent
and in particular we have:
$$\{ {\cal{H}} ,L^{-1}\}~=~[L^{-1},R_{PL}(L)]$$

\noindent
The  above  equation can be trivially but usefully written in terms of
the new dynamical variable $\Lambda\equiv L^{-1}$, taking the  form:

$$\dot \Lambda = [\Lambda, R_{PL}(\Lambda^{-1})] =
[\Lambda, R_{PL}(\nabla_\Lambda {\rm{ln}}det  \Lambda)]\eqno(3.4).$$

\noindent
Thus we have the following {\it{Proposition}}:

``When the variables $(q,p)$ evolve according to the Hamiltonian (1.1),
the ``Flaschka variables" $(A,B)$, defined through (3.3, 3.1b) evolve
according to the Hamiltonian ${\rm{ln}}det  \Lambda$".

\noindent
In this sense, {\it{the PL lattice is just one  of the commuting flows of
the (finite, nonperiodic) Toda hierarchy}}.

Encouraged by the above property, we have further exploited the connection
with the Toda  system to derive an integrable discretisation of the PL
system, i.e. a symplectic map preserving the conserved quantities of the
continuous-time model. Like in the Toda system, the  basic  tool is
the factorisation of the
Lax matrix [12, 15]. Taking care of the natural ordering,
and assuming  for  the sake of simplicity the $+$ sign in $(3.1a,b)$
 it can be written as:

$$L ~=~\sum_j p_jE_{jj}~+~\sum_{k>j}{\sqrt{p_jp_k}}{\rm{exp}}[{\nu \over 2}
(q_k-q_j)] E_{jk}~+~\sum_{k<j}{\sqrt{p_jp_k}}{\rm{exp}}[{\nu \over 2}(q_j-q_k)]
E_{jk}\eqno(3.5)$$
\noindent

It is then possible to find out explicitly (and uniquely, up to signs)
an upper triangular matrix ${\cal{U}}(L)$:

$${\cal{U}}(L)\equiv \sum_{j}\sum_{k\ge j}u_{jk}E_{jk}\eqno(3.6a)$$

\noindent
such that:

$$L~=~{\cal{U}}(L)\times[{\cal{U}}(L)]^t\eqno(3.6b)$$

\noindent
where the superscript $t$ denotes transposition. In fact one finds:

$$u_{jk}^2~=~p_j[{\rm{exp}}[\nu (q_k-q_j)]-{\rm{exp}}[\nu (q_{k+1}-q_j)]
\eqno(3.6c)$$

\noindent
The above expression holds for $1\le k \le N$, provided we set
$q_{N+1}=-\infty$.

\noindent
The  new (or ``Miura transformed") $L$, say $\tilde L$, is  then given
by:

$$\tilde L~=~[{\cal{U}}(L)]^t\times{\cal{U}}(L)\eqno(3.7a)$$

\noindent
which yields for the entries $u_{jk}$ (3.6a) the alternative expression:
$$u_{jk}^2~=~\tilde p_k[{\rm{exp}}[\nu (\tilde q_k-\tilde q_j)]-
{\rm{exp}}[\nu (\tilde q_k-\tilde q_{j-1})]\eqno(3.7b)$$

\noindent
Identifiying (3.6c) and (3.7b) we get, after some trivial manipulation:

$${\tilde p_k \over {\rm{exp}}[\nu(q_k-\tilde q_k)]-{\rm{exp}}
[\nu(q_{k+1}-\tilde q_k)]}~=~{p_j\over {\rm{exp}}[\nu(q_j-\tilde q_j)]-
{\rm{exp}}
[\nu(q_j-\tilde q_{j-1})]}\eqno(3.8)$$

\noindent
The r.h.s. and the l.h.s. of (3.8) must then be independent of the label,
i.e. they have  to be equal to  the same constant, say $\beta$. Therefore,
eq.(3.8) is in fact equivalent to the  canonical transformation:
\noindent
$${\tilde p_k~ =~\beta( {\rm{exp}}[\nu(q_k-\tilde q_k)]-{\rm{exp}}
[\nu(q_{k+1}-\tilde q_k)]})$$

$${p_k~=~\beta( {\rm{exp}}[\nu(q_k-\tilde q_k)]-
{\rm{exp}}
[\nu(q_k-\tilde q_{k-1})]}\eqno(3.9a)$$

\noindent
whose generating function is:

$$S(q,\tilde q)~=~{\beta \over  2}\sum_k({\rm{exp}}[\nu(q_k-\tilde q_k)]-
{\rm{exp}}[\nu(q_{k+1}-\tilde q_k)]\eqno(3.9b)$$

\noindent
The canonical transformation (3.9) can be interpreted as a discrete-time
flow, by setting:

$$(p_j,q_j)=(p_j(n),q_j(n));~(\tilde p_j, \tilde q_j)=
(p_j(n+1),q_j(n+1))$$

\noindent
and written in the following discrete Newton form in terms of
 $q=q(n),~ \tilde q =
q(n+1),~ \hat q=q(n-1):$

$${\rm{exp}}[\nu(\hat q_j - q_j)]-{\rm{exp}}[\nu(\hat q_{j+1}-q_j)]
{}~=~{\rm{exp}}[\nu(q_j-\tilde q_j)]-{\rm{exp}}[\nu(q_j-\tilde q_{j-1})]
\eqno(3.10)$$
\noindent
Moreover, as in the Toda case, such discrete flow can be explicitly
integrated. Namely, if we  denote $L=L(p,q)=L_n,~\tilde L=L(\tilde p, \tilde
q)=
L_{n+1}$, we get [15]:

$$L_n~=~{\cal{U}}(L_0^n)L_0[{\cal{U}}(L_0^n)]^{-1}\eqno(3.11a)$$
\noindent
{}From (3.11a), looking at $n$ as at
a discrete time, we can read off the interpolating hamiltonian
${\cal{F}}(L)$ defined
through:

$$L^n={\rm{exp}}[n\nabla_L{\cal{F}}]$$

\noindent
whence it follows:

$${\cal{F}}(L)~=~tr L({\rm{ln}}L-I)\eqno(3.11b)$$

\noindent
In connection with the above result, it might be worthwhile to recall that
interest in integrable time-discretisation has been recently revived  by
a number of stimulating papers [16].

\vskip 1 true cm
\noindent
{\bf{Concluding remarks}}

We would like to summarize here the main results derived in
the present paper.

First of all, we have found the proper $r-$matrix formulation for the
family of hamiltonian systems recently investigated in [8], containing
as a special case the ``peakons" system: we have provided an alternative
proof of complete integrability and have  exhibited a Lax representation
for the various commuting flows. Secondly, we have noticed that the
PL system enjoys the same $r-$matrix as the finite nonperiodic Toda system:
this property led us to identify the PL system with one of the
commuting
flows of the Toda hierarchy and paved the way for the construction
of an integrable time-discretisation.

Several questions are still open. The most natural one  refers perhaps to the
algebraic nature of the hamiltonian system corresponding to the special choice
$\rho=0$ in (2.3) (yielding $ \mu^\prime=0$ in (1.1)),
already shown to be super-integrable in [13]: namely, is the ``$r-$matrix"
pertaining to this special case still a ``good" one? We did not check Jacobi
identity in this special case, but anyway observe that the corresponding
Lax matrix is of rank 2, and thus one of the conditions which guarantee
complete integrability is not fulfilled (in fact, complete integrability
has been proven in [13] in a different way, not relying upon the
Lax representation). A further open problem is the construction of
an integrable discretisation for the whole family of hamiltonian systems,
not only for the PL system: finding an explicit factorisation is
however a rather hard task in the GL case, and so far we have not
succeded. Finally, the existence of a bi-hamiltonian formulation
 is something to be investigated: in this context, the well known results
available for the Toda system may be exploited in the PL case, while
again the GL system needs a separate investigation, as the dynamical
nature of the $r-$matrix makes the  results derived for instance
 in [14] not
immediately applicable.

\vfill\eject
\noindent
{\bf{References}}

\item {[1]} R.Camassa, D.D. Holm, Phys. rev. Lett. 71 (1993), 1661-1664;

\item { } R. Camassa, D.D. Holm, J.M. Hyman, Adv. Appl. Math. 31 (1994), 1-33.

\item {[2]} A.S. Fokas, B. Fuchssteiner, Physica  D4 (1981), 47-66.
\item {[3]} H. Airault, H.P. McKean, J. Moser, Comm. Pure Appl. Math.
      30 (1977), 95-148.

\item {[4]} I.M. Krichever, Funct. Anal. Appl. 12 (1978), 50-61.

\item { } I.M. Krichever, Funct. Anal. Appl. 14 (1980), 282-290.

\item {[5]} O.I. Bogoyavlenski, S.P. Novikov, Funct. Anal. Appl. 13 (1976), 6.

\item {[6]} E. Previato, ``The Calogero-Moser-Krichever System and Elliptic
Boussinesq Solitons", in {\it{Hamiltonian Systems, Transformation Groups and
Spectral Transform Methods}},

\item { } J.Harnad and J.Marsden eds, CRM Momtreal 1990.

\item {[7]} M. Antonowicz, S. Rauch-Wojciechowski,
Phys. Lett. 147A (1990), 455;

\item { } M. Antonowicz, S. Rauch-Woijciechowski, J. Phys. A: Math. Gen 24
(1992),
5043.

\item {[8]} F.Calogero, J.P. Francoise, ``A  Completely Integrable Hamiltonian
       System", J. Math. Phys. (submitted to).

\item {[9]} M. Alber, ``New types of soliton solutions for nonlinear
equations",

\item { } talk delivered at the Conference on {\it{Nonlinear Coherent
Structures in Physics and Biology}}, held in Edinburg, Scotland, July
 10-14, 1995.

\item {[10]} M.A. Semenov-Tianshanski, Funct. Anal. Appl. 17 (1983), 259.

\item {[11]} O. Babelon, C.M.Viallet, Phys. Lett. B 237 (1990), 411-416.

\item {[12]} W.W Symes, Invent. Math. 59 (1980), 13-51.

\item {[13]} F. Calogero, Phys. Lett. A 201 (1995).

\item {[14]} W. Oevel, O. Ragnisco, Physica A 161, (1989), 181-220.

\item {[15]} Yu. B. Suris, ``New Discretisation of  the Toda Lattice", Preprint
 University of Bremen, Centre for Complex Systems and Visualization, 1995.

\item {[16]} F.Nijhoff, G.D. Pang, ``Discrete-time Calogero-Moser model and
lattice KP equation", in {\it{Symmetries  and Integrability of Difference
Equations}}, D. Levi, P. Winternitz  ans  L. Vinet  eds., CRM Montreal 1995;

\item { } F.Nijhoff, O.Ragnisco, V.Kuzsnetsov, ``Integrable time-discretization
of the Rujisenaars-Schneider model", Comm. Math. Phys. (to appear);

\item { } Yu.B. Suris, ``A discrete-time relativistic  Toda lattice",
Preprint
 University of Bremen, Centre for Complex Systems and Visualization, 1995.

\vfil\end